# A Survey on Security Issues in Ad Hoc Routing Protocols and their Mitigation Techniques

**Harshavardhan Kayarkar**
M.G.M's College of Engineering and Technology, Navi Mumbai, India
Email: hjkayarkar@gmail.com

-----------------------------------------------------------------ABSTRACT-----------------------------------------------------------------
**Mobile Ad hoc Networks (MANETS) are transient networks of mobile nodes, connected through wireless links, without any fixed infrastructure or central management. Due to the self-configuring nature of these networks, the topology is highly dynamic. This makes the Ad Hoc Routing Protocols in MANETS highly vulnerable to serious security issues. In this paper, we survey the common security threats and attacks and summarize the solutions suggested in the survey to mitigate these security vulnerabilities.**

-----------------------------------------------------------------------------------------------------------------------------------------------
-----------------------------------------------------------------------------------------------------------------------------------------------

## 1. INTRODUCTION

MANETs are formed by mobile nodes communicating with each other through wireless links without any governing body. These mobile nodes can be Personal Digital Assistants (PDAs), laptops, cell phones that communicate with each other without any fixed infrastructure and central management. Such networks can be used in the battlefield application, in disaster management and in remote areas where establishment and management of fixed network is not possible. These can also be used in the areas where the establishment of fixed infrastructure is very difficult. MANETs can also be used to deploy and coordinate the drones in the battlefield. They are characterized by unreliable communication media where the network topologies change dynamically. Additionally each node is limited by bandwidth, battery and computation power.

Due to the self-configuring nature of networks and lack of infrastructure, the nodes in the MANETs act both as a router as well as a host. As MANETs are self-developing and highly dynamic, some special ad-hoc routing protocols have been developed. Ad hoc routing protocols should have the following properties:

- **Distributed Operation**: The protocol should be distributed. It should not be dependent on any centralized authority. This is beneficial because the nodes can enter and leave the network easily.
- **Loop Free**: For the efficient functioning of the network, the routing protocol should guarantee that the routes are loop free. This avoids the wastage of bandwidth and computing power. Also, delays are reduced if the routes are loop free.
- **Demand based Operation**: To avoid the unnecessary wastage of bandwidth, computing power and battery, the routing protocol should react only when necessary. In other words, the protocol should be reactive.
- **Unidirectional Link Support**: Unidirectional links are formed in the radio environment. The protocol should use these unidirectional links for the optimal performance of the protocols.
- **Security**: Security is an important issue in MANETs. MANETs are susceptible to attacks like spoofing. To guarantee the desired behavior in ad hoc routing protocols some security measures are required. Security can be improved by applying encryption and authentication to the routing protocols.
- **Quality of Service Support**: Quality of Service is an important parameter in the ad hoc routing protocols. The routing protocols should support various QoS. For instance, real time traffic should have low jitter. It should be noted that none of the proposed protocols have all these properties.
- **Multiple routes**: The protocol should have redundant routes, so when one link fails an alternative route can be used without initiating route discovery. Also, buffering multiple routes makes the protocol resistant to frequent topology changes.
- **Power conservation**: The nodes that form the ad hoc network have very limited resources. One such important resource which is limited is the battery power. The protocols should conserve the battery power of the mobile devices. They should switch to power saving or standby mode when not in use.

Some of the commonly used ad hoc routing protocols in MANETs are Destination-Sequenced Distance Vector (DSDV) [1] routing, Dynamic Source Routing (DSR) [1], [2], Ad-hoc On-demand Distance Vector (AODV) routing [6], Optimized Link State Routing (OLSR) [58] and Zone Routing Protocol (ZRP) [6]. A brief summary of these ad hoc protocols is presented below.

**Summary of different Ad-hoc Routing Protocols**

- Divecha et al. [1] have carried out the performance analysis of DSDV and DSR protocols and compared their performances with different mobility models. They concluded that the routing protocols are specific to particular mobility models.
- Ramesh et al. [2] have proposed a method to reduce the end to end delay in the multi-path routing protocol by proposing a congestion aware multi-path DSR protocol. It enhances the performance of DSR protocol in congested network. The proposed protocol was compared with ordinary DSR protocol and the results show that the proposed scheme greatly reduces the end to end delay and improves the overhead.
- Williams and Camp [3] have presented a comprehensive comparison between different broadcasting schemes used in MANETs. In their paper, they categorized different broadcasting schemes and compared them through simulations, which established various network failures under different conditions like bandwidth consumption, dynamic topologies, and battery consumption. They have also proposed some protocols extension that can adapt to the changing network conditions and improve the functioning of the broadcasting scheme.
- Adibi and Agnew [4] in their paper presented a survey on different versions of DSR, pointed out their differences and compared them. The authors have also proposed a multilayer flavored DSR protocol which obtains the information from physical, MAC and network layer and passes this information to the network layer. Then they select the most optimal routing protocol by performing a comparison between the current network condition and the pre-defined closest group of conditions.
- Toubiana et al. [5] performed a comparison of different multipath reactive routing protocols. The authors have compared three node-disjoint multipath routing protocols and two routing protocols based on an Untrusted Node Disjoint (UND) path scheme. The comparison showed that the multipath scheme limited the exerted performance degradation compared to UND. The comparison also showed that the multipath scheme incurred more overhead in the safe and secured network.
- Mittal and Kaur [6] compared the performance of DSR, AODV and ZRP protocols. The comparison was performed on various metric and the results show that AODV performed the best in achieving the end to end delay and packet delivery ratio while DSR performed the best with minimum number of hops in comparing TTL based hop count.
- Gowrishankar et al. [58] compared the performance of AODV and OLSR protocols. For a network with static traffic and smaller number of nodes, AODV performs well as it uses fewer amounts of resources. OLSR shows higher efficiency in networks where the node density is high and the traffic pattern is random.

This paper is organized as follows: Section 2 explains some of the common security threats and vulnerabilities in MANETs. Section 3 presents the techniques to mitigate the security threats in MANETs. Section 4 gives the Comparative Analysis of the techniques discussed in Section 3. Section 5 draws the conclusion of this paper.

## 2. SECURITY THREATS AND VULNERABILITIES IN MANETS

Due to the inherent property of the MANETs of being structure-less, dynamic, self-configuring and self-sustaining in nature, there exist some potential loopholes and vulnerabilities in MANETs that can be attacked and exploited by the malicious and undesirable nodes to disrupt the smooth functioning in the network. Some of the common attacks in MANETs are:

- **Impersonation or Spoofing**
  The objective of this attack is to hide the real identity of the attacker. In this attack, the attacker assumes the identity of a more trusted node in the network. By doing this the other nodes include this malicious node in their routing path and the malicious node can then disrupt the normal functioning of the network without being noticed.
- **Black-hole Attack**
  The purpose of this attack is to increase the congestion in network. In this attack the malicious node does not forward any packets forwarded to it, instead drops them all. Due to this attack the packets forwarded by the nodes do not reach their intended destination and the congestion in the network escalates due to retransmissions.
- **Sink-hole Attack**
  The idea of the attacker in this attack is to attract all the network traffic towards itself. The attacker executes this attack by making the neighboring nodes believe that the shortest path to the destination is through it. This attack causes the other nodes to relay all the traffic through the malicious node so that the attacker can modify, fabricate or just listen to the received packets.
- **Wormhole Attack**
  The main aim of the wormhole attack is to replay the packet on the other side of the network. This attack is executed by two nodes colluding to form a wormhole. The attacker on one side make the nodes believe that distance to the destination is just one hop, when it is greater than one hop. This causes the attacker to attract all the traffic from one side of the network and relay it through the wormhole; the attacker on the other side replays the same packet. By doing this the attacker can drop the packets or obtain any service illegally.

- **Sleep Deprivation**
  The goal of the attacker in this attack is to keep the target node constantly busy. This attack is initiated by flooding the network with routing traffic and thereby making the node consume all of the computing and battery power. This attack forces the targeted node in consuming the battery, network bandwidth and computing power by spurious requests for existent or non-existent destination nodes, so that it cannot process the legitimate requests.

- **Rushing Attack**
  The purpose of this attack is to include the malicious node in the routing path. During the route discovery phase the RREQs are forwarded by the malicious nodes to the neighbors of the target node. These RREQs are quick to reach the neighboring nodes. When a neighboring node receives this hurried RREQ from the attacker, it will not forward any request originated from the source node that initiated route discovery. By executing this attack the attacker includes itself in the routing table and can then tamper with the packet.

- **Location Disclosure**
  The location disclosure attack intends to target the privacy requirements of the ad-hoc network. In this attack the attacker, by doing traffic analysis or using simple monitoring, approaches and finds the location of the destination node in the network. By knowing the intermediary nodes the attacker can find the node of concern and gain the information about the structure and the topology of the network.

- **Routing Table Poisoning**
  The intention of the attacker in this attack is to corrupt the routing table. The routing protocols maintain the routing tables to find a route to the destination and forward the packet to the anticipated node. In this attack, the malicious nodes generate and send the fabricated traffic into the network or modify the legitimate messages from other nodes. An alternative way to execute this attack is by broadcasting a RREQ with higher sequence number in the network resulting in the valid packets with lower sequence number to get rejected. This attack causes the routing tables to create wrong entries and store the corrupt invalid information in the routing tables of the participating nodes.

- **Route Fabrication**
  The objective of this attack is to gain illegal access to the packets or to initiate packet dropping in network. In this attack, an attacker hinders with the normal routing procedures. This attack is executed by changing the routing messages or by inserting false routing messages. Due to the fabrication of routing information the packets are routed to non-existent nodes or they can be forwarded to a malicious node. It also results in the delay of the packets and bandwidth wastage. Routing fabrication also leads to Denial of Service (DoS) attacks.

- **Denial Of Service (DoS) Attack and Flooding**
  The aim of this attack is to cripple the smooth functioning of the network. This attack is accomplished by continually sending packets into the network causing the targeted node in the network to process them and keep them occupied resulting in the crashing of that node. By executing this attack, the attacker keeps the targeted node busy in processing its fabricated packets and depriving the legitimate RREQs to be dropped. This attack can cause the network infrastructure to collapse.

- **Routing Table Overflow**
  The principle of this attack is to consume the routing table buffer. The attacker sends false and engineered traffic into the network. It modifies the packets and includes the routes to non-existent nodes. This attack overwhelms the routing table buffer by storing false routing information in the routing table. This blocks the processing of legitimate routes and storing them since the routing table runs out of space. This kind of attack can be particularly executed and severely affect the DSR protocol as it stores the route to every node resulting in the exhaustion of the route cache.

## 3. TECHNIQUES TO MITIGATE VARIOUS SECURITY ATTACKS IN MANETS

In this section we will be presenting a survey on different techniques that are proposed to enhance and fortify the ad hoc routing protocols against various security loopholes and vulnerabilities in the ad hoc networks.

### 3.1 Solution Using Trust Values

In this subsection, various trust value based solutions have been discussed. These solutions mitigate various security vulnerabilities and enhance the existing ad hoc routing protocols.

Pirzada and McDonald [7] present a method to improve the DSR protocol. They propose the method of deploying trust gateways to reinforce the DSR protocol. In this method, the number of malicious nodes in the network is identified and with the use of the trust gateways, they are avoided in the future exchange of data packets.

In the Trusted Dynamic Source Routing [8] model by Yong et al., trust among nodes is calculated using a combination of direct and indirect trust. When the trust value of a node declines so much that it falls below a threshold, it is then added to a blacklist. The packets from the blacklisted nodes are not forwarded.

Dhurandher and Mehra [9] have employed a message trust based solution to the multipath routing scenario. In this proposed solution each node is initially given a zero trust value indicating an unknown trust level. Based on the behavior of the nodes the assigned trust value is either incremented or decremented. Trust values may be positive, negative or zero, indicating known, malicious, or unknown behavior.

Mangrulkar and Atique [10] presented a scheme that enhances the AODV protocol by adding an extra field in the RREQ called Trust Value. The initial trust value is assigned by the source when it broadcasts RREQ packet. When it receives RREP from the destination it increments the trust value of all the nodes that fall on the route of destination. By adding this extra field the source selects a valid route that has higher trust value rather than selecting the shortest or the longest route. This avoids the disruption of the network as most of the attacks are coordinated on the shortest route to the destination.

Hallani and Shahrestani [11] used fuzzy logic for trust assessment in mobile ad hoc networks. The trusted nodes are then used in selecting the most reliable communication path.

Sen [12] has proposed an approach for detection of malicious packet dropping attack on MANETs. The process depends on the trust of each node which in turn is computed by analyzing the packet forwarding behavior of the nodes.

Buchegger and Boudec proposed CONFIDANT [13] protocol that uses a trust manager to identify misbehaving nodes. Such nodes are prevented from sending and forwarding RREQ and RREP.

Halim et al. employ Agent-based trusted solution for securing the DSR protocol [14]. In the proposed model the authors use a multi-agent system (MAS) consisting of two types which collaborate with each other to achieve the required task, that is, the monitoring agent (MOA) and routing agent (ROA). The MOA monitors its hosting node behavior in the routing process and then calculates the trust value for the node. ROA is in-charge of using this trust value and locating the most reliable path to the destination.

Ngai and Lyu [15] propose a method to provide secure and distributed authentication service in ad hoc networks. The proposed solution uses secure public key authentication service based on the trust model and network to prevent nodes from gaining false public keys of other nodes.

The method proposed by Yu et al. [16] prevents against Byzantine attacks in MANETs using encrypted messages and route redundancy. The route is selected based on trustworthiness of the neighboring modes.

Vasudevan and Sanyal [17] proposed an approach to securely transfer data packets in MANETs. The suggested technique uses the properties of the polynomials in an encryption-less algorithm. The primary step in the proposed scheme is to obfuscate the data in such a way that it is impossible to gather any information without acquiring the complete obfuscated data. The next step of the algorithm is to split the obfuscated data into various pieces where each data piece appears like puzzle of the jigsaw. These individual pieces are then sent over disjoint paths to the intended destination. This authentication method also provides the guarantee that every packet is authentic.

### 3.2 Wormhole Detection Method

This section presents the solution to prevent and detect wormhole in ad hoc networks.

Lee et al. have proposed a solution to mitigate the wormhole attack in MANETs [18]. Here the wormhole is collectively detected by route identification. Each node maintains its neighbor's information, thus identifying the route that is suffering from the wormhole.

Sharma and Trivedi [19] proposed a method to defend against the wormhole attack. In the proposed solution the authors use digital signatures to prevent against the wormhole attack. Whenever a node wants to send a packet it initiates RREQ. Along with RREQ it also sends its digital signature. The nodes in the network verify this digital signature with the one stored in their database and if there is match, they confirm that the RREQ is from a legitimate source. The malicious node replaying the RREQ either has a signature of other node or does not have any and hence is identified and isolated from further transmission.

### 3.3 Intrusion Detection Systems (IDS)

This section provides different Intrusion Detection and Prevention Systems.

Negar et al. proposed an Intrusion Detection System (IDS) [20] based on the interaction between the user and the kernel processes. A feature list is created to distinguish between the normal and anomalous behavior. They introduce a new function to the Linux Kernel called the Wrapper Module to log initial data to prepare the intended feature list. For the classification of the input vectors SVM neural network is applied in the proposed scheme. The authors tried to improve the accuracy, training time and testing time as compared to other systems. Wei and Wu [21] present a method that combines feature extraction and SVM (Support Vector Machine) model to improve the classification accuracy and minimize the detection time. In the feature extraction using CEGA (Classification Ensemble by Genetic Algorithms), the fitness of the individual is calculated by classification rate and conditional entropy. The SVM model on the other hand is processed simultaneously that finds the optimal feature subset. The proposed method is effective in intrusion detection. Visumathi and Shunmuganathan [22] proposed architecture for Intrusion Detection System that uses SVM classifiers. The authors presented a survey of the existing intrusion detection systems.

Penvaand and Bringas [23] proposed a method that detects the misuse and the anomalies in the network. The authors use Bayesian networks to learn the misuse and anomalies and use this knowledge to further detect other known and unknown attacks.

Saravanakumar et al. [24] tackle the issue of complexity and throughput that are evident in the current Intrusion Detection Systems (IDS). They compare various IDS systems that use different algorithms to detect the intrusions. The authors also propose a scheme that uses a combination of Artificial Neural Network algorithms to design an IDS which enables faster convergence and delivers better performance. Shun and Malki [25] have proposed a scheme to improve Intrusion Detection System (IDS) in ad hoc networks. This scheme uses neural networks to predict and detect the attacks on network. The authors use the concept of feed forward neural networks with back propagation training. By training the proposed IDS system with the attacks the presented scheme greatly enhances the IDS system and detects the known and unknown attacks correctly and with higher probability. Chavan et al. [26] developed an Intrusion Detection System (IDS) using Fuzzy Inference System and Artificial Neural Networks. The proposed IDS system is trained by creating a signature pattern database. It uses protocol analysis and Neuro-fuzzy learning method.

Dal et al. have proposed an Intrusion Detection System method to secure the network [27]. The proposed method applies Genetic Algorithm with Artificial Immune System (AIS) to develop an Intrusion Detection. In this paper the authors evolve a Primary Response following the concept of memory cells which is prevalent in Natural Immune System. These memory cells enable faster detection of already encountered attacks. These cells are highly random in nature and they are dependent on the evolution of the detectors. This grants greater immunity from anomalies and attacks. An extensive survey by Dasgupta et al. [28] focusses on the recent advances in Artificial Immune System. Yang et al. [29] use a similar method in AIS to enhance the Intrusion Detection System. The method uses antibody concentration to evaluate the danger level of the intrusion in the network. Hosseinpour et al. [30] described a method to improve the detection performance and accuracy of IDS system. The authors propose a distributed multilayered framework to improve the detection and efficiency of IDS. The genetic algorithm proposed in the paper enhances the secondary immune response of the system. Jie et al. [31] proposed a method for signal detection using AIS for anomalistic signal detection in an electromagnetic environment.

Saboori et al. [32] devised a method to improve the intrusion detection. The authors use Apriori Algorithm to detect an anomaly in the system. This algorithm predicts a novel attack and generates a set of real-time rules for the firewall. The algorithm functions by extracting the correlation relationships among the large data sets. Nikolova and Jecheva [33] proposed an anomaly based Intrusion Detection System (IDS). In the proposed scheme the authors use data mining techniques like classification trees to describe the normal activity of the system. To detect the intrusion in the system, similarity coefficients are used that compare the similarity between the normal behavior and the observed behavior. Depending on the measured similarity a decision is made if the system is under attack or not.

Karim [34] illustrated the success and the contributions of Computer Intelligence in the Network Intrusion Detection. The author explains the importance of computer intelligence in the intrusion detection of clustering, feature selection, and anomaly detection.

Abraham et al. [35] have proposed an Intrusion Detection System. In this proposed method they compare the performance of their fuzzy rule based classifiers for IDS with similar performance obtained from the decision tree, support vector mechanisms and linear genetic programming. Soft Computing (SC) based IDS is used to develop a light weight and more accurate IDS. Toosi et al. [36] presented a method to classify the normal and abnormal behavior in network. The authors proposed Adaptive Neuro Fuzzy Inference system to categorize into normal and suspicious behavior and detect intrusion. Binary and multi-classifiers are used to perform this job.

Faysel and Haque [37] provided a comprehensive survey on different Intrusion Detection Systems. They also explore the Intrusion Prevention Systems that were proposed in the recent years. The authors present the limitations and shortcomings of some of the Intrusion Detection and Prevention System in use.

Trivedi et al. [38] proposed a Semi Distributed Reputation-based IDS method for MANETs. This method employs the concepts of redemption and fading. Redemption system allocates and maintains the standing of different nodes whereas the path manager performs minor path management.

Banerjee et al. [39] proposed an Ant Colony based IDS to keep track of the intruder trails. The proposed method works in conjunction along with the default learning based detection systems and provides higher level of security to the sensor networks.

Platos et al. [40] presented a method to strengthen the network. The authors propose an improvement over the earlier work on Non-negative Matrix Factorization approach. A GPU implementation is done to improve the speed and accuracy and overall performance of the system.

Sodani et al. [41] presented different firewalls and filtration techniques, their drawbacks and shortcomings. The authors also propose a method collaborating with the Application Layer Filtering (ALF) and logging and learning to develop a stronger and efficient filtering technique to prevent attacks on the network.

Trivedi et al. [42] proposed an Intrusion Detection System. The method defines a reputation that is assigned to every node in the network. Each node monitors the behavior of its next-hop neighbor through promiscuous mode. The reports of the action of the nodes are submitted to the reputation manager for updating the reputation value. Whenever a node crosses a predefined threshold it is declared as malicious and a warning message is sent only to the immediate neighbors. Each node also contains an avoid list that contains a list of malicious nodes and no further communication is done through these nodes.

### 3.4 Black-hole detection and prevention

This section presents the solution to black-hole detection and prevention method.

Wang and Shi [43] proposed a scheme to secure DSR protocol based on request sequence number. The scheme uses creditable routing information which is formed and based on the acknowledgements received by the source from the destination. The information of the routing table is centered on the source node of the RREQ and is divided according to the trust value which in turn decides the routing path of the RREQ packets. This ensures that every node has valid information and black-hole attack can be prevented.

Cai et al. [44] introduced a method to detect black-hole and gray-hole attacks in ad hoc network. The authors have proposed a path-based method that overhears the next hop's actions. As the scheme does not send out control messages it saves the system resources. To lower the false positive rate under high network overload, a collision rate reporting system is established in the MAC layer. This adaptive threshold approach decreases the false positive rates.

Khan et al. [45] introduced a solution to prevent against packet dropping in MANETs. In the proposed solution the authors suggest a two-fold approach. In the first step the identification of the malicious node is made and in the second step the isolation of that node is done. Whenever intermediary nodes receive the packet for the destination from the source they send acknowledgement to the source. If the source does not receive any acknowledgement from a particular intermediate node even after retransmitting the packet, then it concludes that the node is misbehaving and makes an entry of that node. This information is forwarded to all the other nodes in network and the transmission activity is stopped and the malicious node is isolated.

Moradiya and Sampalli [46] devised a method to prevent and detect routing intrusion in OLSR protocol. The authors propose a mechanism that verifies the control messages sent by the intruder and detect the intrusion thereby preventing routing intrusion. The HNA message in the OLSR protocol is verified before updating the kernel routing table. When the HNA messages are validated only then the kernel routing table is updated, otherwise an alert is generated and the intrusion is detected. The proposed mechanism can be used to detect and prevent the black-hole attack in the network.

### 3.5 Sink-hole Detection and Prevention Method

Culpepper and Tseng [47] have introduced Sinkhole detection system to combat sink-hole in DSR protocol. In this system there are three variables: Sequence Number Discontinuity, Previous Image Ratio and Route Add Ratio. All these three variables tell if there is a sinkhole present in the network and if their values are high, low and high as compared to a predetermined value.

Sheela et al. [48] have presented a method to defend against the sink-hole attack in wireless sensor networks. In the proposed method the authors use what they call as the mobile agent programs. These agents travel to each and every node in the ad hoc network, collect the information and update the routing table of the nodes with the latest information. This mechanism causes every node to be aware of all other nodes in the network and this allows it to ignore the bogus information from the malicious node, trying to launch the sink-hole attack.

Thumthawatworn et al. [49] proposed a method to detect sink-hole. In the proposed method the detection of sink-hole is achieved by applying trust-based algorithm to all the nodes in the network. Separation of suspicious nodes from normal nodes is achieved by using the corresponding weights and threshold values.

### 3.6 Credibility Management and Routing Test

Pengwei and Zhenqiang [50] have proposed a method that enhances the security of the AODV protocol. In this proposed method the authors enhance the security of the AODV by declaring a neighbor table. This neighbor table contains three fields: the neighbor IP, expired time and Credit value. Initially, the credit value of the node is set to 1. When a node sends the data packet in the network it stores a copy of that packet in its buffer. Then, when other nodes rebroadcast the packet and if the listened packet is same as the packet stored in the buffer of the listening node's buffer, it increments the credit value of the neighboring node in the neighbor buffer table, otherwise decrements the credit value by some factor.

### 3.7 Link Cache Updating

DSR protocol maintains cache to keep the redundant routes. Over time the routes in cache become stale. To avoid this problem Yu [51] proposed a scheme to update the cache and keep the fresh routing information. In the proposed method the author defines a cache table and defines a distributed cache algorithm. Necessary information for cache updates is stored by each node in its cache table. Whenever a link failure is detected in the network, the algorithm spreads this information to all the nodes in the network link and updates their related cache table. This process is completed in a distributed manner. The proposed method does not depend on any ad hoc parameters of the network and hence is fully adaptive to the topology changes.

### 3.8 Flooding Attack Prevention Technique

Different solutions to prevent the Flooding Denial of Service (DoS) attack have been discussed in this section.

Kataria et al. [52] have proposed a scheme to control the flooding of fake route requests in ad-hoc networks. To control the flooding of fake route requests and to ensure fairness to genuine RREQs, a parameter known as RREQ_RATELIMIT is considered, it is fairly distributed among all the participating nodes. This limited bandwidth which is allotted to each node restricts the number of RREQs injected in the network and processed by each node.

Jia et al. [53] proposed a method to prevent the Denial of Service (DoS) attack in multi-path communication in Mobile Ad hoc Networks (MANETs). A capability message has been defined that is exchanged by each node. This enables them to maintain a global view of the overall throughput of the flow in the network and dynamically adjust to local constraints to prevent a DoS attack. The presented method alleviates DOS attack by regulating the end to end traffic transmitted over the network.

Ahmad et al. [54] have proposed a method to reduce the flooding in the network due RREPs from different nodes. The author uses DSR protocol for comparison. The flooding of RREPs in the network is achieved by delaying the RREP to the source so that the route with the least hop count will reach the source earlier than other RREPs. The delaying also avoids the problem of overwhelming the source with unwanted RREPs.

Sasson et al. [55] have proposed a method to limit the unnecessary flooding of packets in the network. Plain flooding of packets in the network causes unnecessary congestion and packet collision and bandwidth consumption. The authors have explored the phase transition phenomenon observed in percolation and random graphs as basis to define probabilistic flooding algorithm. They have compared this algorithm with plain flooding and have shown that the use of probabilistic flooding algorithm in routing protocols may benefit MANETs.

### 3.9 Data Hiding Technique

Dey et al. [56] introduced a data hiding technique. This technique is based on the decomposition of number in sum of prime numbers. This generates a different set of bit-planes which is suitable for embedding. This enables embedding of secret messages in higher bit planes without causing any distortion. A better stego-image quality is achieved in reliable and secured manner, guaranteeing efficient retrieval of data. A comparison between the classical Least Significant Bit (LSB) method, the Fibonacci LSB data-hiding technique and the proposed scheme was carried. It was observed that stego-image hidden was indistinguishable from the original cover-image. This idea can be extended to increase security in ad hoc networks.

### 3.10 First Fast Second Reliable Method (FFSR)

Zhai et al. [57] have presented a method that improves the reliability and efficiency of AODV protocol. In this paper, they propose an idea called "First Fast Second Reliable (FFSR)" that ensures the transfer of data packages in the shortest and the most reliable mode. The routing metric in this method is based on the cognition of each node and all the nodes in cognitive ad hoc network collect information about the whole network periodically.

### 3.11 Multi-Factor Authentication Techniques

In [59], [60] authors have proposed a Multi-factor security authentication method for wireless payment. This idea can very easily be extended to provide better security in ad hoc networks protocol. In [61] this technique has been used to prevent impersonation in ad hoc networks.

## 4 COMPARATIVE ANALYSIS OF DIFFERENT TECHNIQUES

In this section we provide a quantitative comparison between the different techniques discussed in the previous section to mitigate various attacks.

### 4.1 Detection of Malicious Nodes on Trust Value Technique

In this subsection, we compare different trust value techniques to detect and avoid the malicious nodes.

The solution proposed in [7] uses Trust Gateways to identify the malicious nodes whereas the method proposed in [8] employs direct and indirect trust as compared to simple Trust Gateways discussed earlier. The advantage of this method is that it maintains a blacklist that contains the list of misbehaved nodes. The nodes from this list are excluded while forming the routing information. The method proposed in [9] uses positive, negative and zero trust values to identify the known, malicious or unknown behavior of the nodes in the network. The technique proposed by [12] employs a rather different solution to calculate and assess the trust values. It uses fuzzy logic to establish the trust levels of the node. The scheme suggested in [13] employs watchdog and path rater concept. In this method the watchdog detects the misbehaving nodes and the path rater makes sure that these nodes are not included in the forwarding routes. From simulations results, it has been observed that this protocol performs normally even when a large percentage of the nodes in the network are malicious.

From the above comparison of various trust based techniques we infer that the trust values are utilized to detect the malicious nodes in the network thereby avoiding them and selecting optimal communication path.

### 4.2 Comparison of techniques to mitigate Wormhole attack

This subsection compares various techniques to detect and prevent the wormhole attack.

The method proposed in [18] employs a cooperative method to detect a wormhole. Each node in this scheme maintains its neighbor's information. This technique detects the wormhole attack in the network. On the other hand the technique suggested in [19] uses digital signatures to prevent the wormhole attack. Each node, when receiving the RREQ, cross-checks the digital signature in the packet with the one stored in its routing table. If it is legitimate then it forwards the packet otherwise it informs other nodes that the previous node that forwarded the packet is a malicious node, thus preventing the wormhole attack.

### 4.3 Comparison of different IDS Techniques

We compare different IDS techniques discussed in the previous section.

The solutions proposed in [20], [21], [22] use SVM model to design Intrusion Detection System. The method suggested in [20] uses a wrapper module along with SVM network to improve the accuracy, testing time of the IDS. On the other hand CEGA [21] is employed together with SVM model to make the IDS system more effective. The solution proposed in [22] employs plain SVM classifiers in the architecture of the IDS system. A different method is suggested in [23]. Bayesian networks are used, that learn from the misuse of the network and the anomalies and detects other known and unknown attacks.

The techniques suggested in [24], [25], [26], [35], [36] employ Neural Networks and Neuro-Fuzzy Inference systems to enhance the IDS as compared to the SVM methods proposed in the previous paragraph.

In [32] and [33], Data Mining techniques are used to detect the intrusions. The method proposed in [33] uses Classification Trees to distinguish between the normal and the abnormal behavior whereas the scheme suggested in [32] uses Apriori Algorithm to detect anomalies in the system. It predicts the attack and then generates a set of real-time rules for the firewall to defend against the related attacks.

The schemes stated in [27], [29], [31] employ Genetic Algorithm and Artificial Immune System to develop an Intrusion Detection System. The method in [27] uses memory cells to evolve the primary response whereas anomalistic signal detection [31] is used in electromagnetic environment. Antibody concentration [29] is used to evaluate the danger level of the intrusion in the network.

### 4.4 Comparison of different Black-hole detection Methods

In this subsection we provide the comparison of various techniques discussed in the preceding section.

The method suggested in [44] uses path based approach that overhears the next hop's actions. This scheme does not send out control messages thus reducing the overhead. On the other hand, the technique proposed in [46] verifies the control messages sent by the intruder.

A two-fold scheme [45] is employed that prevents the black-hole attack. In this scheme the acknowledgements are sent to the source by the nodes when they receive the packet. When the source does not receive the acknowledgment, it infers that a malicious node is present in the network.

### 4.5 Comparison of Sinkhole Detection Methods

In this subsection we compare the various techniques discussed to mitigate sinkhole detection attack.

In [47] a sinkhole detection scheme is presented. In this scheme three variables are used that detect if sinkhole is present in the network or not. Alternatively, mobile agents [48] are used in wireless sensor networks to detect the sinkhole. In this method the mobile agents travel to each node collecting the information and making each node aware of others, thus preventing the sinkhole attack.

A trust-based method [50] is used that compares with a predefined threshold to detect the sinkhole.

### 4.6 Comparison of techniques to prevent the Flooding Attack

Here we compare various techniques that are applied to prevent the flooding attack.

In [52] a parameter known as RREQ_RATELIMIT is used to put a limit on the number of RREQs introduced in the network. This prevents the flooding of RREQs. A solution presented in [54] delays the RREPs in the network so that the source does not get overwhelmed with unnecessary RREPs, thus preventing the flooding attack.

A different approach is chosen in [55] as compared to the earlier schemes in [52] and [53]. In this method phase transition phenomenon, discussed in percolation and random graphs, is used to define the probabilistic flooding algorithm.

TABLE 1 lists the summary of the security attacks and the techniques to mitigate them in brief.

Table 1: A summary of various Security Threats and techniques to mitigate them

| Security Attack | A brief description of the Security Attack | Techniques proposed to mitigate the attacks | Ad Hoc Routing Protocol employed |
|---|---|---|---|
| 1. Wormhole Attack | The attack is executed by colluding of nodes. The attacker replays the packet from one side of the network to the other side. | 1. In [18] the authors use a cooperative approach among the distributed nodes to detect and mitigate the wormhole attack.<br><br>2. The method proposed in [19] detects and prevents and the wormhole by implementing digital signatures. | 1. The proposed scheme [18] can be incorporated with any of the ad hoc routing protocols.<br><br>2. The authors in [19] have combined their method in AODV and DSR protocols and implemented them. |
| 2. Black-hole Attack | The attacker drops all the packets forwarded to it in this attack. | 1. The method in [43] uses creditable routing table to detect the black-holes and eradicate them.<br><br>2. The authors in [44] employ a path-based approach to detect and mitigate the black-hole attack.<br><br>3. In [45] a two-stage approach is adopted. In stage one, the detection of the malicious nodes is performed and in the next stage the isolation of the malicious node is carried out.<br><br>4. The authors in [46] verify the control messages sent by the intruder and detect if the black-hole attack is executed. | 1. The authors in [43] use DSR protocol to test their method.<br><br>2. In [44] the method is combined with the DSR protocol for testing purposes.<br><br>3. The author's method in [45] can be integrated with any ad hoc routing protocol like AODV, DSR etc.<br><br>4. OLSR protocol is employed in [46] to assess the proposed method. |
| 3. Sink-hole Attack | The attacker attracts the entire network traffic towards itself and can modify or fabricate the received packets. | 1. In [47] a method that uses three variables namely Sequence Number, Route Add Ratio and Previous Image Ratio is implemented to prevent, detect and mitigate the sinkhole attack.<br><br>2. The authors use a scheme that employs mobile agents to detect and mitigate the sink-hole attack [48].<br><br>3. A trust based algorithm [49] is implemented to alleviate sink-hole attack | 1. In [47] DSR protocol is employed to mitigate the sinkhole attack.<br><br>2. The method [48] is integrated with the AODV protocol to test the scheme.<br><br>3. The author suggest [49] using on-demand multi-path routing protocol to implement the proposed method. |
| 4. Flooding Attack | In this attack the attacker floods the network with fabricated traffic to cripple the smooth and efficient functioning of the network. | 1. The authors in [52] proposed a method that controls the flooding attack by introducing a variable known as RREQ_RATELIMIT which restricts the number of packets sent into the network.<br>2. In [53] a Capability Based messages are exchanged by every node to maintain the global view of the throughput thereby failing the flooding attack.<br>3. In [54] the flooding attack is controlled by delaying the RREP from destination to the source so that the number of messages sent in the network is reduced.<br>4. A probabilistic flooding algorithm in [55] to mitigate the flooding attack. Phase transition phenomenon is explored to define the algorithm. | 1. The authors [52] use AODV protocol to implement their proposed scheme.<br>2. The proposed scheme in [53] uses AOMDV protocol for simulation and testing.<br>3. The author employs DSR protocol for the implementation of the proposed scheme [54].<br>4. The algorithm suggested in [55] can be integrated with any ad hoc routing protocol. |
| 5. Route Fabrication | False routing messages or fabricated messages are inserted into the network by the attacker in this attack. | 1. In [11] the authors use fuzzy logic, a soft computing method to establish a quantifiable trust value among the nodes of the network. This approach prevents the route fabrication attack. | 1. The suggested scheme [11] is integrated with the AODV protocol for the implementation. |
| 6. Impersonation/Spoofing | In this attack the attacker assumes the identity of another node in the network. | 1. The authors in this method [15] use secure public key authentication based on the trust model. | 1. The proposed method [15] can be combined with any ad hoc routing protocol. |

## 5. CONCLUSION

In this paper we surveyed various protocols for MANETs, discussed various security vulnerabilities to mitigate the attacks and presented the performance analysis of several routing protocols.

In the Introduction section we discussed about MANETs, listed their advantages and how they are formed. We also discussed different ad hoc routing protocols, explained the working of each and provided a table that lists the ad hoc routing protocols and the properties exhibited by each of them. In the next section we listed various security vulnerabilities and threats that are encountered in the MANETs. We explained each of the security threats and the effects they cause on the ad hoc networks. In the last section we discussed several techniques to mitigate the security threats and attacks listed in the previous section. We categorized the solutions based on various techniques.

We also provided a comparative analysis of different techniques that are employed to mitigate diverse security issues and presented a table that summarizes the entire section 4.

**REFERENCES**

[1] Bhavyesh Divecha, Ajith Abraham, Crina Grosan. Sugata Sanyal, "Analysis of Dynamic Source Routing and Destination-Sequenced Distance-Vector Protocols for Different Mobility models", *First Asia International Conference on Modeling and Simulation*, AMS2007. March, 27-30, 2007, Phuket, Thailand. Publisher: IEEE Press, pp. 224-229.

[2] V. Ramesh, P. Subbaiah, N. Sandeep Chaitanya, K. Sangeetha Supriya, "Performance Comparison of Congestion Aware Multi-Path Routing (with Load Balancing) and Ordinary DSR", *2010 IEEE 4th International Conference on Internet Multimedia Services Architecture and Application(IMSAA),* Dec. 15-17, 2010, pp.1-5.

[3] Brad Williams, Tracy Camp, "Comparison of Broadcasting Techniques for Mobile Ad Hoc Networks", *MOBIHOC'02,* June 9-11, 2002, EPFL, Lausanne, Switzerland, pp. 194-205.

[4] S. Adibi, G.B. Agnew, "Multi-layer flavored dynamic source routing in mobile ad-hoc networks", *IET Communications*, 2008, Vol. 2, No. 5, pp. 690–707.

[5] Vincent Toubiana, Houda Labiod, Laurent Reynaud and Yvon Gourhant, "Performance Comparison of Multipath Reactive Ad hoc Routing Protocols", *IEEE 19th International Symposium on Personal, Indoor and Mobile Radio Communications (PIMRC 2008),* Sept. 15-18, 2008, pp.1-6.

[6] Shaily Mittal, Prabhjot Kaur, "Performance Comparison of AODV, DSR and ZRP Routing Protocols in MANETS", *International Conference on, Advances in Computing, Control, and Telecommunication Technologies, 2009. ACT '09*, Dec. 28-29, 2009, pp.165-168.

[7] Asad Amir Pirzada, Chris McDonald, "Deploying Trust Gateways to Reinforce Dynamic Source Routing", *2005 3rd IEEE International Conference on Industrial Informatics, (INDIN '05),* Aug. 10-12, 2005, pp. 779- 784.

[8] CHENG Yong, HUANG Chuanhe, SHI Wenming, "Trusted Dynamic Source Routing Protocol", *Wireless Communications, International Conference on Networking and Mobile Computing, WiCom2007*, Sept. 21-25, 2007, pp.1632-1636.

[9] Sanjay K. Dhurandher, Vijeta Mehra, "Multi-path and Message Trust-Based Secure Routing in Ad Hoc Networks", *International Conference on Advances in Computing, Control, and Telecommunication Technologies, ACT '09.*, Dec. 28-29, 2009, pp.189-194.

[10] R. S. Mangrulkar, Mohammad Atique, "Trust Based Secured Ad hoc on Demand Distance Vector Routing Protocol for Mobile Ad Hoc Network", *2010 Sixth International Conference on Wireless Communication and Sensor Networks (WCSN),* Dec. 15-19, 2010, pp.1-4.

[11] H. Hallani, S.A. Shahrestani, "Trust Assessment in Wireless Ad-hoc Networks", *Wireless Days, 2008 (WD '08). 1st IFIP*, *Dubai,* Nov. 24-27, 2008, pp.1-5.

[12] Jaydip Sen, "A Distributed Trust and Reputation Framework for Mobile Ad Hoc Networks", *Proceedings of the 3rd International Conference on Network Security and Applications*, Chennai, India, 2010, pp. 538- 537.

[13] Sonja Buchegger, Jean-Yves Le Boudec, "Performance Analysis of the CONFIDANT Protocol (Cooperation Of Nodes: Fairness In Dynamic Ad hoc NeTworks)", *Proceedings of IEEE/ACM Workshop on Mobile Ad Hoc Networking and Computing (MobiHOC'02),* June 9-11, 2002, EPFL Lausanne, Switzerland, pp. 226-236.

[14] Islam Tharwat A. Halim, Hossam M. Fahmy, Ayman, M. Bahaa El-Din, Mohamed H. El-Shafey, "Agent-based Trusted On-Demand Routing Protocol for Mobile Ad-hoc Networks", *2010 4th International Conference on Network and System Security (NSS)* Sept. 1-3, 2010, pp. 255-262.

[15] Edith C.H. Ngai, Michael R. Lyu, "Trust- and Clustering-Based Authentication Services in Mobile Ad Hoc Networks", *Proceedings of 24th International Conference on Distributed Computing Systems Workshops, 2004*, March 23-24, 2004, pp. 582- 587.

[16] Ming Yu, Mengchu Zhou, Wei Sou, "A Secure Routing Protocol against Byzantine Attacks for MANETs in Adversarial Environments", *IEEE Transactions on Vehicular Technology,* vol.58, no.1, Jan.2009, pp.449-460.

[17] R. Vasudevan, Sugata Sanyal, "A Novel Multipath Approach to Security in Mobile and Ad Hoc Networks (MANETs)", *Proceedings of International Conference on Computers and Devices for Communication (CODEC'04)*, Kolkata, India, December, 2004, pp. CAN_0412_CO_F_1 to CAN_0412_CO_F_4.

[18] Gunhee Lee, Dong-kyoo Kim, Jungtaek Seo, "An Approach to Mitigate Wormhole Attack in Wireless Ad Hoc Networks", *International Conference on Information Security and Assurance (ISA 2008)*, April 24-26, 2008, pp.220-225.

[19] Pallavi Sharma, Aditya Trivedi, "An Approach to Defend Against Wormhole Attack in Ad Hoc Network Using Digital Signature", *2011 IEEE 3rd International Conference on Communication Software and Networks (ICCSN)*, May 27-29 2011, pp.307-311.

[20] Almassian Negar, Azmi Reza, Berenji Sarah, "AIDSLK: An Anomaly Based Intrusion Detection System in Linux Kernel", Information Systems, Technology and Management *Communications in Computer and Information Science*, 2009, Publisher: Springer Berlin Heidelberg, pp. 232-243.

[21] Yuxin Wei, Muqing Wu, "Intrusion detection technology based on CEGA-SVM," *Third International Conference on Security and Privacy in Communications Networks and the Workshops, (SecureComm 2007)*, Sept. 17-21, 2007, pp.244-249.

[22] J. Vishumathi, K.L Shunmuganathan, "A computational intelligence for evaluation of intrusion detection system", *Indian J. of Science and Technology*, Jan. 2011, Issue 1, Vol. 4, pp. 40-45.

[23] Penva, Y. K, Bringas, P. G., "Integrating network misuse and anomaly prevention," *6th IEEE International Conference on Industrial Informatics, 2008. INDIN 2008*, July 13-16, 2008, pp.586-591.

[24] S. Saravanakumar, Umamaheshwari, D. Jayalakshmi, R. Sugumar, "Development and implementation of artificial neural networks for intrusion detection in computer network", *Int. Journal of Computer Science and Network Security*. 2010. vol. 10, no. 7, pp. 271-275.

[25] Jimmy Shun and Heidar A. Malki, "Network Intrusion Detection System Using Neural Networks", *Fourth International Conference on Natural Computation,(ICNC '08)*, vol.5, Oct. 18-20, 2008, pp.242-246.

[26] Sampada Chavan, Khusbu Shah, Neha Dave, Sanghamitra Mukherjee, Ajith Abraham, Sugata Sanyal, "Adaptive Neuro-Fuzzy Intrusion Detection Systems", *IEEE International Conference on Information Technology: Coding andComputing,* 2004 (ITCC '04), Proceedings of ITCC 2004, Vol. 1, April, 2004, Las Vegas, Nevada, pp. 70-74.

[27] Divyata Dal, Siby Abraham, Ajith Abraham, Sugata Sanyal, Mukund Sanglikar, "Evolution Induced Secondary Immunity: An Artificial Immune System based Intrusion Detection System", *7th International Conference on Computer Information Systems and Industrial Management Applications, (CISIM '08)*, June 26-28, 2008, pp.65-70

[28] D. Dasgupta, S. Yu, F. Nino, "Recent Advances in Artificial Immune Systems: Models and Applications", *Applied Soft Computing*, Elsevier, Vol. 11, March, 2011, pp.1574-1587.

[29] Jin Yang, Yi Liu, JianJun Wang, JianDong Zhang, Bin Li, "Dynamical Immunological Surveillance for Network Danger Evaluation Model," *5th International Conference on Wireless Communications, Networking and Mobile Computing (WiCom '09), Beijing, China,* Sept. 24-26, 2009, pp.1-4.

[30] F. Hosseinpour, K. Abu Bakar, A. Hatami Hardoroudi, A. Farhang Dareshur, "Design of a new distributed model for Intrusion Detection System based on Artificial Immune System," *2010 6th International Conference on Advanced Information Management and Service (IMS), Seoul, Korea,* Nov. 30-Dec. 2, 2010, pp.378-383.

[31] MA Jie, SHI Ying-chun, ZHONG Zi-fa, LIU Xiang, "An Anomalistic Electromagnetism Signal Detection Model Based on Artificial Immune System," *2010 International Conference on Communications and Intelligence Information Security (ICCIIS),NanNing, China*, Oct. 13-14, 2010,pp.256-260.

[32] E. Saboori, S. Parsazad, Y. Sanatkhani, "Automatic firewall rules generator for anomaly detection systems with Apriori algorithm," *2010 3rd International Conference on Advanced Computer Theory and Engineering (ICACTE),Chengdu, China,V*ol.6 , Aug. 20-22, 2010, pp.V6-57-V6-60.

[33] Evgeniya Nikolova, Veselina Jecheva, "Some similarity coefficients and application of data mining techniques to the anomaly-based IDS", *Telecommunication Systems*, December, 2010, Publisher: Springer Netherlands, pp. 1-9.

[34] Asim Karim, "Computational Intelligence for Network Intrusion Detection: Recent Contributions and Security", *Computational Intelligence and Security, International Conference, CIS 2005, Xi an, China,* December 15-19, 2005, Proceedings, Part I. Volume 3801 of *Lecture Notes in Computer Science*, pp. 170-175.

[35] Ajith Abraham, Ravi Jain, Sugata Sanyal, Sang Yong Han, "SCIDS: A Soft Computing Intrusion Detection System",*6th International Workshop on Distributed Computing (IWDC-2004)*, Springer Verlag, Germany, Lecture Notes in Computer Science, Vol. 3326., 2004, pp. 252-257.

[36] A. N. Toosi, M. Kahani, R. Monsefi, "Network Intrusion Detection based on Neuro-fuzzy classification," *International Conference on Computing & Informatics, (ICOCI '06), Kuala Lumpur, Malaysia*, June 6-8, 2006,pp.1-5.

[37] Mohammad A. Faysel, Syed S. Haque, "Towards Cyber Defense: Research in Intrusion Detection and Intrusion Prevention Systems", *IJCSNS International Journal of Computer Science and Network Security*, Vol.10 No.7, July 2010, pp. 316-325.

[38] Animesh K Trivedi, Rajan Arora, Rishi Kapoor, Sudip Sanyal, Sugata Sanyal, "A Semi-distributed Reputation-based Intrusion Detection System for Mobile Ad hoc Networks", Journal of Information Assurance and Security (JIAS), Volume 1,Issue 4, December, 2006, pp. 265-274.

[39] S. Banerjee, C. Grosan, A. Abraham, P. K. Mahanti, "Intrusion detection in sensor networks using emotional ants," *Proceedings of 5th International Conference on Intelligent Systems Design and Applications, (ISDA '05), Wroclaw, Poland*, Sept. 8-10, 2005, pp. 344- 349.

[40] J. Platos, P. Kromer, V. Snasel, A. Abraham; "Scaling IDS construction based on Non-negative Matrix factorization using GPU computing," *Sixth International Conference on Information Assurance and Security (IAS), Atlanta, USA*, Aug. 23-25, 2010, pp.86-91.

[41] Rajesh Sodani, A.K Goyal, Basant Singh Rathore, "MLF: A Technology beyond ALF for Network Security", *2011 International Conference on Communication Systems and Network Technologies (CSNT), Katra, Jammu, India*, June 3-5, 2011, pp. 58-61.

[42] Animesh Trivedi, Rishi Kapoor, Rajan Arora, Sudip Sanyal, Sugata Sanyal, "RISM - Reputation Based Intrusion Detection System for Mobile Ad hoc Networks", *3rd International Conference on Computers and Devices for Communication (CODEC-06),* Institute of Radio Physics and Electronics, University of Calcutta, December 18-20, 2006, pp. 234-237.

[43] Japing Wang, Haoshan Shi, "A Secure DSR Protocol Based on the Request Sequence-Number", *5th International Conference on Wireless Communications, Networking and Mobile Computing, 2009 (WiCom '09), Beijing, China,* Sept. 24-26, 2009, pp. 1-4.

[44] Jiwen CAI, Ping YI, Jialin CHEN, Zhiyang WANG, Ning LIU, "An Adaptive Approach to Detecting Black and Gray Hole Attacks in Ad Hoc Network", *2010 24th IEEE International Conference on Advanced Information Networking and Applications (AINA),Perth, Australia,* April 20-23, 2010, pp.775-780,.

[45] Muhammad Zeshan, Shoab A. Khan, Ahmad Raza Cheema, Attique Ahmed, "Adding Security against Packet Dropping Attack in Mobile Ad hoc Networks", *International Seminar on Future Information Technology and Management Engineering, (FITME '08), Leicestershire, UK,* Nov. 20. 2008, pp.568-572.

[46] Pradeep Moradiya, Srinivas Sampalli, "Detection and Prevention of Routing Intrusions in Mobile Ad Hoc Networks", *2010 IEEE/IFIP 8th International Conference on Embedded and Ubiquitous Computing (EUC), Hong Kong,* Dec. 11-13, 2010,pp.542-547.

[47] Benjamin J. Culpepper, H. Chris Tseng, "Sinkhole Intrusion Indicators in DSR MANETs", *Proceedings of First International Conference on Broadband Networks(BroadNets 2004),San Jose, USA,* Oct. 25-29, 2004, pp. 681- 688.

[48] D. Sheela, Naveen Kumar. C, G. Mahadevan, "A Non-Cryptographic Method of Sink Hole Attack Detection n Wireless Sensor Networks", *2011 International Conference on Recent Trends in Information Technology (ICRTIT),Chennai, India,* June 3-5, 2011, pp.527-532.

[49] Thanachai Thumthawatworn, Tapanan Yeophantong, Punthep Sirikriengkrai, "Adaptive Sinkhole Detection on Wireless Ad Hoc Networks", *Proceedings of IEEE Aerospace Conference, 2006,Big Sky, Montana, USA*, 4-11 March 2006,pp.1-10.

[50] Li Pengwei, Xu Zhenqiang, "Security Enhancement of AODV against Internal Attacks",*2010 2nd International Conference on Information Science and Engineering (ICISE), Hangzhou, China,* Dec. 4-6, 2010,pp.584-586.

[51] Xin Yu, "Distributed Cache Updating for the Dynamic Source Routing Protocol", *IEEE Transactions on Mobile Computing,*Vol.5, no.6, June 2006, pp. 609- 626.

[52] Jayesh Kataria, P.S. Dhekne, Sugata Sanyal, "A Scheme to Control Flooding of Fake Route Requests in Ad-hoc Networks", *3rd International Conference on Computers and Devices for Communication (CODEC-06)* Institute of Radio Physics and Electronics, University of Calcutta, December 18-20, 2006, pp. 198-201.

[53] Quan Jia, Kun Sun, Angelos Stavrou, "CapMan: Capability-based Defense against Multi-Path Denial of Service (DoS) Attacks in MANET", *Proceedings of 20th International Conference on Computer Communications and Networks (ICCCN),Maui, HI, USA, 2011,* July 31-August 4, 2011, pp.1-6.

[54] Shakeel Ahmad, Irfan Awaan, Athar Waqqas, Bashir Ahmad, "Performance Analysis of DSR & Extended DSR Protocols", *Second Asia International Conference on Modeling & Simulation,(AICMS 08), Kuala Lumpur,* May 13-15, 2008, pp.191-196.

[55] Yoav Sasson, David Cavin, Andre Schiper, "Probabilistic Broadcast for Flooding in Wireless Mobile Ad hoc Networks", *2003 IEEE Wireless Communications and Networking, (WCNC 2003), New Orleans, LA, USA,*vol.2, vol.2, March 202003, pp.1124-1130.

[56] Sandipan Dey, Ajith Abraham, Sugata Sanyal," An LSB Data Hiding Technique Using Prime Numbers", *Third International Conference on Intelligent Information Hiding and Multimedia Signal Processing, (IIHMSP 2007),Kaohsiung City, Taiwan, IEEE Computer Society press, USA,*vol.2, Nov. 26-28, 2007, pp.473-476.

[57] Zhenhui Zhai, Yong Zhang, Mei Song, Guangquan Chen, "A Reliable and Adaptive AODV Protocol based on Cognitive Routing for Ad Hoc Networks"*, 2010 The 12th International Conference on Advanced Communication Technology (ICACT),Gangwon-Do, Korea,* vol.2, Feb. 7-10, 2010, pp.1307-1310,.

[58] S. Gowrishankar, T.G. Basavaraju, M. Singh, Subir Kumar Sarkar, "Scenario based Performance Analysis of AODV and OLSR in Mobile Ad Hoc Networks", *Proceedings of the 24th South East Asia*

*Regional Computer Conference*, Bangkok, Thailand, Nov. 18-19, 2007, pp. 120-128.

[59] Ayu Tiwari, Sudip Sanyal, Ajith Abraham, Sugata Sanyal, "A Multifactor Security Protocol For Wireless Payment-Secure Web Authentication using Mobile Devices", *IADIS International Conference, Applied Computing 2007, Salamanca, Spain*. Feb. 17-20, 2007, pp. 160-167.

[60] Sugata Sanyal, Ayu Tiwari, Sudip Sanyal, "A Multifactor Secure Authentication System for Wireless Payment", *Emergent Web Intelligence: Advanced Information Retrieval Book Series: Advanced Information and Knowledge Processing*, First Edition, 2010, Chapter 13, pp. 341-369.

[61] D. Glynos, P. Kotzanikolaou, C. Douligeris; "Preventing impersonation attacks in MANET with multi-factor authentication," *Third International Symposium on Modeling and Optimization in Mobile, Ad Hoc, and Wireless Networks, 2005(WIOPT 2005), Trentino, Italy,* April 3-7, 2005, pp. 59- 64.

## Author(s) Biography

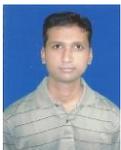

Harshavardhan Kayarkar is currently in his final year of Bachelor of Engineering from M.G.M's College of Engineering and Technology, University of Mumbai, Maharashtra, India. He is planning to pursue higher studies in Networking and Network Security.